 \documentclass[final,3p,times]{elsarticle}

\usepackage{amssymb}
\usepackage{graphicx}
\usepackage{booktabs}

\begin{document}

\begin{frontmatter}


\title{Behavior patterns of online users and the effect on information filtering}
\author[]{Cheng-Jun Zhang}
\author[]{An Zeng\corref{cor2}}
\ead{an.zeng@unifr.ch} \cortext[cor2]{Corresponding author}

\address{Department of Physics, University of Fribourg, Chemin du Mus\'{e}e 3, CH-1700 Fribourg, Switzerland\\}

\begin{abstract}
Understanding the structure and evolution of web-based user-object bipartite networks is an important task since they play a fundamental role in online information filtering. In this paper, we focus on investigating the patterns of online users' behavior and the effect on recommendation process. Empirical analysis on the e-commercial systems show that users have significant taste diversity and their interests for niche items highly overlap. Additionally, recommendation process are investigated on both the real networks and the reshuffled networks in which real users' behavior patterns can be gradually destroyed. Our results shows that the performance of personalized recommendation methods is strongly related to the real network structure. Detail study on each item shows that recommendation accuracy for hot items is almost maximum and quite robust to the reshuffling process. However, niche items cannot be accurately recommended after removing users' behavior patterns. Our work also is meaningful in practical sense since it reveals an effective direction to improve the accuracy and the robustness of the existing recommender systems.

\end{abstract}

\begin{keyword}
bipartite networks, reshuffling process, information filtering
\end{keyword}

\end{frontmatter}

\section{Introduction}
Complex networks have been studied intensively for more than a decade. The rapid development of network science has greatly helped us to understand and model real systems~\cite{PR175}. So far, many systems have been described by networks including the transportation system~\cite{PNAS7794}, neural system~\cite{Complexity56}, social system~\cite{PRL108701}, power grid~\cite{PRE025103} and so on. Some other systems coupled by two different elements are modeled by the bipartite networks. For example, the online commercial systems~\cite{EPL48006} and the scientific collaborative systems~\cite{PRE056103} are well represented by such networks.

With the help of the network structure, many novel methods have been proposed to improve the function of real systems. The online commercial system is a good example. Nowadays, we can simply order books, movies, clothes from the online retailer even at home. However, like a coin has both sides, internet also brings us overabundant information so that we always have too many candidate products to compare. In order to solve the problem, many recommendation algorithms such as collaborative filtering~\cite{ACM61}, content-based analysis~\cite{LNCS325},
spectral analysis~\cite{PRL248701} and iterative self-consistent refinement~\cite{EPL58007} were developed to filter irrelevant information. Recently, some physical dynamics on the bipartite networks, including mass diffusion~\cite{PRE046115}  and heat
conduction process~\cite{PRL154301}, have been applied to design recommendation algorithms. Hybrid of these so-called network-based inference methods (NBI) is shown to have significant improvement in both recommendation accuracy and item diversity compared to the traditional methods~\cite{PNAS4511,PRE066119}.

When studying the recommendation, most of the previous works are devoted to improve the performance of the recommendation algorithms by examining their methods on some standard datasets~\cite{ACM5}. However, the network structure properties will inevitably affect the recommendation process~\cite{IJC138}. For example, given a recommendation method, its the performance would change from one network to another. Actually, how much do the recommendation method rely on typical data are still unclear. To answer such question, it is useful to investigate the users' online behaviors in different real systems. Previous works have found that people's behavior are far different from random and obey certain predictable rules~\cite{NP818}. Therefore, the users' online behavior will emerge some typical statistical patterns on the network structure. Consequently, these patterns will influence the recommendation process based on the networks.

In this paper, we focus on understanding online users' statistical behavior pattern and the related effect on information filtering. We will compare the real bipartite networks with the randomized counterpart networks (i. e. the reshuffled networks) in which real users' behavior pattern are destroyed. Actually, some specific properties of the real networks has been discovered by the comparison to the reshuffled networks such as the loop distribution~\cite{PRL118701,PRE046121}, rich club~\cite{NP110,PRL168702}, community structure~\cite{PNAS7821}, assortative~\cite{PRL208701} and motifs~\cite{Science824}. Here, we find that online users have significant taste diversity and their interests for niche items highly overlap. Additionally, recommendation process are investigated on both the real networks and the reshuffled networks. We find that the performance of popularity-based recommendation methods don't rely on the real network structure while the performance of personalized recommendation methods is strongly related to it. Detail study on the personalized methods indicate that recommendation accuracy for hot items is almost maximum and quite robust to the reshuffling process. On the contrary, niche items cannot be accurately recommended without real users' behavior properties. Moreover, our work is meaningful in practical aspect since it reveals an effective direction to improve the accuracy and the robustness of the existing recommender systems.

\section{Statistical behavior pattern of online users}

\begin{figure}
  \center
  \includegraphics[width=9cm]{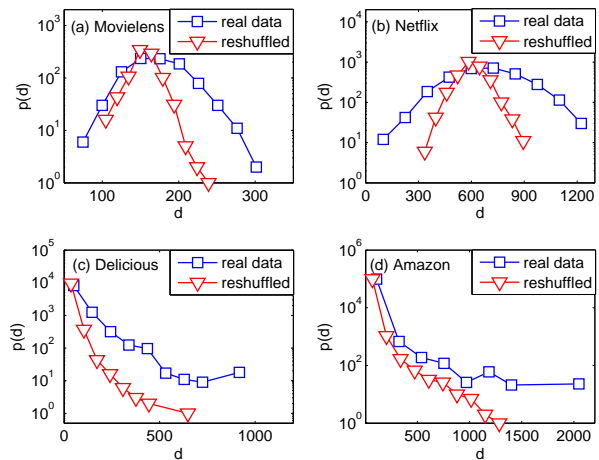}\\
  \caption{(Color online) The distribution $P(d)$ in real systems where $d$ is the average degree of selected items for each user.}
\end{figure}

\begin{table}[!htb]
 \tabcolsep=4pt
\begin{center}
\caption{Properties of the used datasets}
\begin{tabular}{lccclccc}
\hline
  network  &Users &Items  &Links &Sparsity\\
\hline
  Movielens &$943$ &$1,682$ &$82,520$ &$5.20\cdot10^{-2}$\\
  Netflix &$3,000$ &$3,000$ &$197,248$ &$2.19\cdot10^{-2}$\\
  Delicious &$10,000$ &$232,657$ &$1,233,997$ &$5.30\cdot10^{-4}$\\
  Amazon &$99,622$ &$645,056$ &$2,036,091$ &$3.17\cdot10^{-5}$\\
 \hline\label{table}
\end{tabular}
\end{center}
\end{table}

In this paper, the datasets that we will use are the subsets of data obtained from four online systems: Movielens (http://www.grouplens.com/), Netflix (http://www.netflixprize.com/), Delicious (http://www.delicious.com/) and Amazon (http://www.amazon.com/). These data are random samplings of the whole records of user activities in these websites, the descriptions of data are given in Table I.

To investigate users' behavior pattern, we will compare the real bipartite networks with the reshuffled networks. In each step of the reshuffling process, we first randomly pick two links from the real network, for example, one is from
user $i$ to item $\alpha$ and the other is from user $j$ to item $\beta$ (throughout this paper we
use Greek and Latin letters, respectively, for object- and user-related indices). Then we
rewire these two links by $i$ to $\beta$ and $j$ to $\alpha$. Hence, the
degree of the users and items would not be changed by this reshuffling
process while the links in this reshuffled networks are
randomized. Denoting $T$ as the reshuffling times and $L$ as the total links in the networks, we fix $T/L=3$ in the following analysis.

After the reshuffling process, users' degree and items' degree are preserved while the correlation between users and items are destroyed. To begin our comparison, we focus on the average degree of users' selected items. Suppose a user $i$ selects $m$ items with degree $k_{\alpha}$ ($\alpha=1,2,...,m$), we calculate the average degree of the items that he/she selected as $d_{i}=\frac{\sum_{\alpha=1}^{m}k_{\alpha}}{m}$. Actually, the distribution of $d$ reflects the taste diversity of the users. When all the users prefer the same type of items, users' $d$ will be the same to each other. Consequently, the distribution of $d$ will be extremely narrow. On the contrary, the distribution of $d$ will be quite flat if all the users seek for different items. We then compare the distribution $P(d)$ in real networks and their reshuffled networks. As shown in fig. 1, $P(d)$ in real networks indeed are much boarder than that in the reshuffled networks. Obvious, users have obvious taste diversity in real systems.

Secondly, for each user we study the inter-similarity among all his/her selected items. The similarity of two items is calculated by the common neighbor here~\cite{CN1971}. Suppose a user $i$ selects $m$ items and the similarity between item $\alpha$ and $\beta$ is denoted as $s_{\alpha\beta}$, the inter-similarity among all these $m$ items can be obtained by $\widetilde{S}_{i}=\frac{2\sum^{m}_{\alpha=2}\sum^{\alpha}_{\beta=1}s_{\alpha\beta}}{m(m-1)}$. In fact, $\widetilde{S}$ indicates the taste diversity for each single user. Specifically, when a user always select for the same kind of items, $\widetilde{S}$ for him/her will be high. On the other hand, if the interest of a user changes from time to time, his/her $\widetilde{S}$ will be very low. As shown in fig. 2, compared to the reshuffled networks, the \emph{inactive users} (i.e. users with small degree) in real systems show a higher $\widetilde{S}$ while the \emph{active users} (i.e. users with large degree) are with lower $\widetilde{S}$. Actually, since the inactive users in real networks do not have much experience in seeking for their own interested objects, they tend to conservatively choose several most popular objects. Hence, their selected items are very similar. On the contrary, active users in real systems are more likely to explore and try different kinds of unpopular objects. Therefore, their selected items are with low $\widetilde{S}$.

\begin{figure}
  \center
 \includegraphics[width=9cm]{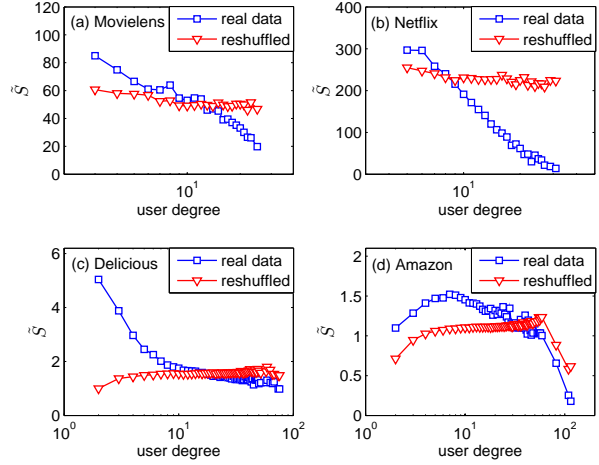}\\
  \caption{(Color online) The inter-similarity $\widetilde{S}$ among all the selected items for each user vs user's degree. For a given $x$, its corresponding $\widetilde{S}$ is obtained by averaging over all the items whose degrees are in the range of $[a(x^2-x),a(x^2+x)]$, where $a$ is chosen as $\frac{1}{2}log5$ for a better illustration.}
\end{figure}

\begin{figure}
  \center
 \includegraphics[width=9cm]{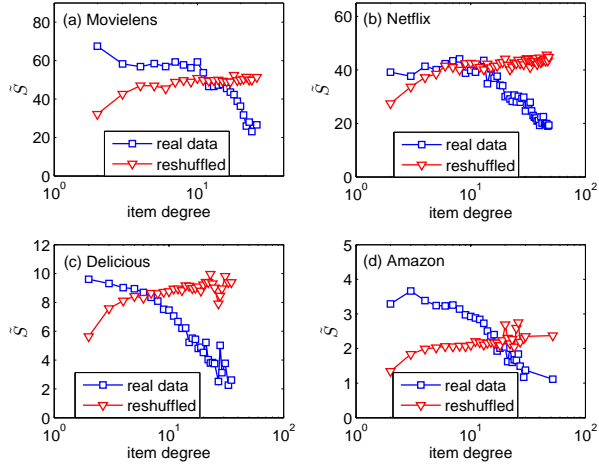}\\
  \caption{(Color online) The inter-similarity $\widetilde{S}$ among all the selecting users for each item vs item's degree. The $\widetilde{S}$ is averaged by the same process in fig.2.}
\end{figure}

Similarly, for each item we investigate the inter-similarity among all the users who selected it. Assume a item $\alpha$ is chosen by $n$ users and the similarity between user $i$ and $j$ is denoted as $s_{ij}$, the inter-similarity among all these users can be calculated by $\widetilde{S}_{\alpha}=\frac{2\sum^{n}_{i=2}\sum^{i}_{j=1}s_{ij}}{n(n-1)}$. Actually, $\widetilde{S}$ reflects whether a specific item is selected by the same group of users. In fig.3, we study $\widetilde{S}_{\alpha}$ in the real networks and the reshuffled networks. For \emph{hot items} (i.e. item with large degree), their selectors in real networks have a lower $\widetilde{S}$ than those in the reshuffled networks. However, the selectors of \emph{niche items} (i.e. item with small degree) enjoy a higher inter-similarity in the real networks than those in the reshuffled networks. As we know, the personalized recommendation systems generally filter relevant information by cooperating the history of similar users, the overlap of users' interests for niche items is very meaningful. It makes the limited historical information for these niche items valuable for the recommendation systems to refer to. In next section, we will detailedly investigate how these users' online behavior patterns affect the recommendation process.

\section{The effect on information filtering}

In order to reveal the effect of users' online behavior patterns on information filtering, we investigate the recommendation process on both the real networks and the reshuffled networks in which real users' behavior patterns are destroied. We consider four conventional recommendation algorithms including mass diffusion (MD), heat conduction (HC), collaborative filtering (CF), popularity-based (PR) methods. We will study how the recommendation performance is influenced when we gradually remove users' real behavior patterns.

We first briefly describe these algorithms.
Consider a system of $N$ users and $M$ items represented by a bipartite network
with adjacency matrix $A$,
where the element $a_{i\alpha}=1$ if user $i$ has collected object $\alpha$, and $a_{i\alpha}=0$ otherwise. For a target user $i$,
the MD algorithm starts by
assigning one unit of resources to objects collected by $i$, and redistributes the resource through the user-item network.
We denote the vector $\textbf{f}$ as the initial resources on items where $f_{\alpha}$ is the resource possessed by object $\alpha$. The redistribution is represented by $\widetilde{\textbf{f}}=W\textbf{f}$, where
\begin{equation}
W_{\alpha\beta}=\frac{1}{k_{\beta}}\sum\limits_{l=1}^{N}\frac{a_{l\alpha}a_{l\beta}}{k_{l}},
\end{equation}
is the diffusion matrix, with $k_{\beta}=\sum^{N}_{i=1}a_{i\beta}$ and $k_{l}=\sum^{M}_{\gamma}a_{l\gamma}$ denoting the degree of object $\beta$ and user $l$ respectively~\cite{PRE046115}. Technically, recommendations for a given user $i$ are obtained by setting
the initial resource vector $\textbf{f}^{i}$ in accordance with the objects the user has
already collected, that is, by setting $f^{i}_{\alpha}=a_{i\alpha}$. The resulting recommendation
list of uncollected objects is then sorted according to $\widetilde{f}^{i}_{\alpha}$ in descending order. Physically, the diffusion is equivalent to a three-step random walk starting with $k_i$ units of resources on the target user $i$. The recommendation score of an item is taken to be the amount of resources on it after the diffusion. The scores for objects that user $i$
have already collected are set to $0$. The recommendation
list for user $i$ is generated by ranking all his/her uncollected
objects in descending order of their final resources.

The HC algorithm works similar to the MD algorithm, the only difference is the diffusion matrix is calculated as
\begin{equation}
W_{\alpha\beta}=\frac{1}{k_{\alpha}}\sum\limits_{l=1}^{N}\frac{a_{l\alpha}a_{l\beta}}{k_{l}}.
\end{equation}
Physically, the temperature of an object is considered to be the average temperature of its nearest neighborhood, i.e. its connected users. The higher the temperature of an item, the higher its recommendation score~\cite{PRL154301}.

The CF algorithms provide recommendations based on user or item similarities. Here, we consider the item-based CF which has been
successfully applied to many online applications such as
Amazon (one of the largest online product retailers). In the item-based CF method, the recommendation score of an item is evaluated based on its similarity with the collected items of the target user. The final recommendation score for each item can be written as
\begin{equation}
\widetilde{f}^{i}_{\alpha}=\sum^{M}_{\beta=1}s_{\alpha\beta}a_{i\beta}.
\end{equation}
where $s_{\alpha\beta}$ is the similarity between item $\alpha$ and $\beta$~\cite{ACM5}. The measure of similarities used in CF is subject to definition.
Here we simply define the similarity as the number of common neighbors in the bipartite networks.

The PR algorithms is very simple and commonly used in many websites. In this method, the recommendation score for each item is proportional to its popularity.

Actually, the difference of these recommendation methods has been studied in detail in ref.~\cite{arxiv1106}. In order to further understand these methods, we calculate the total recommendation score for each item as $F_{\alpha}=\sum_{i=1}^{N}\widetilde{f}^{i}_{\alpha}$. The result is shown in fig. 4. In statistical sense, the MD, CF and PR methods assign high recommendation score to the high degree items. In HC method, the items with low degree are generally with high recommendation score. Therefore, the MD, CF and PR methods tend to recommend the popular items while the HC method inclines to recommend unpopular items.

\begin{figure}
  \center
 \includegraphics[width=9cm]{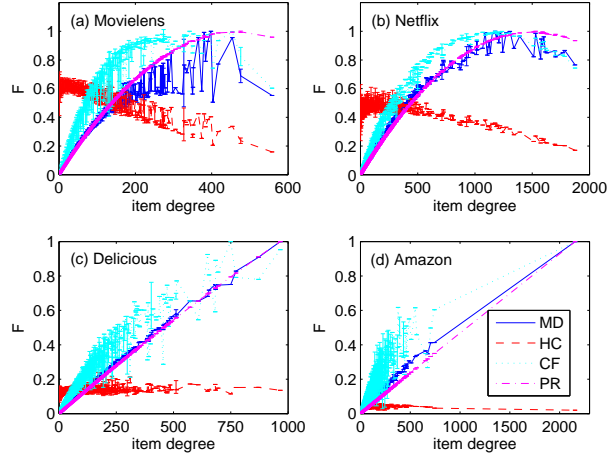}\\
  \caption{(Color online) The total recommendation score $F$ vs item degree in different recommendation systems. The maximum $F$ for each method has been scaled to $1$.}
\end{figure}

We then apply all these methods to the real networks and their reshuffled networks to see how users' real behavior patterns affect the recommendation.
Similar to previous work~\cite{PNAS4511}, to test the recommendation result we randomly remove $10\%$ of the links (the probe set denoted as $E^{P}$). We then apply the algorithms to
the remainder (the training set denoted as $E^{T}$) to produce a recommendation list for each user.

In order to measure the accuracy of the recommendation result, we make use of the ranking score index~\cite{PRE046115}. For a
target user, the recommender system will return a ranking list of all his uncollected
objects to him according to the recommendation scores. For each hidden user-object relation (i.e., the link in probe set), we
measure the rank of the object in the recommendation list of this user. For example, if
there are $1000$ uncollected objects for user $i$, and object $\alpha$ is at $10$th place, we say the
position of this object is $10/1000$, denoted by $RS_{i\alpha}=0.01$. A successful recommendation result is expected to
highly recommend the items in the probe set, and thus leading to small ranking score. Averaging over
all the hidden user-object relations, we obtain the mean value of ranking score to evaluate the recommendation accuracy, namely
\begin{equation}
<RS>=\frac{1}{|E^{P}|}\sum_{i\alpha\in E^{P}} RS_{i\alpha},
\end{equation}
where $i\alpha$ denotes the probe link connecting user $i$ and object $\alpha$. Clearly, the smaller
the ranking score, the higher the algorithm's accuracy, and vice versa.

In fig. 5, we report how the ranking score of different recommendation methods will be influenced when we gradually remove real users' behavior patterns. The results show that the ranking score of PR is hardly affected by the reshuffling process. It is reasonable because the PR method doesn't rely on the detail bipartite network structure and gives the recommendation score for each item simply according to its popularity. On the contrary, the personalized recommendation such as the MD, HC and CF methods are influenced. Obviously, the ranking score of HC method increases most significantly when we reshuffle the networks. In fact, the HC method is considered as an effective method to enhance recommendation diversity by mainly predicting users' preference for niche items. Therefore, the result implies that without the real correlation between users and items, only the information of degree is insufficient for the recommendation systems accurately providing a diverse recommendation. More specifically, as we discussed in the previous section, users' interests for niche items highly overlaps in real systems. Hence, the recommendation systems can predict target user's potential niche items by cooperating the information from his/her similar users. However, in the reshuffled networks users' interests for niche items only slightly overlap, so there is little information from the similar users for the recommendation engines to refer to. It finally leads to the serious increment in the ranking score of HC method.

\begin{figure}
  \center
   \includegraphics[width=9cm]{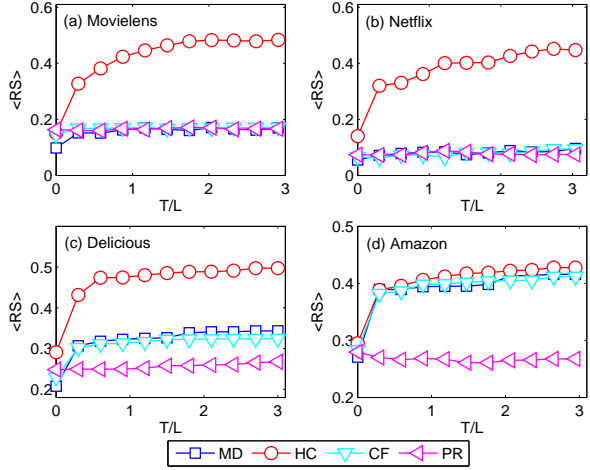}\\
  \caption{(Color online) The ranking score $<RS>$ of different recommendation methods when reshuffling the real networks. $T$ is the reshuffling steps and $L$ is the total links in the networks.}
\end{figure}

As recommendation algorithms which tend to recommend popular items, MD and CF methods are not so sensitive to the reshuffling process as the HC method. In the dense networks like Movielens and Netflix, the ranking scores of MD and CF stay almost unchanged. However, in the sparse networks like Delicious and Amazon, the ranking score of MD and CF methods show an observable increment. In order to see the effect of the reshuffling process on the MD and CF methods in detail, we study the ranking score for each item, namely
\begin{equation}
<RS_{\alpha}>=\frac{1}{|E^{P}_{\alpha}|}\sum_{i\alpha\in E^{P}_{\alpha}} RS_{i\alpha},
\end{equation}
where $E^{P}_{\alpha}$ denotes all the links in the probe set that connect to item $\alpha$. Then we can see the relation between items' degree and their ranking score, the result is reported in fig. 6. In real networks, the hot items enjoy a low ranking score ($<RS>\approx0$) while the niche items are with high ranking score (It can be even higher than the random recommendation whose $<RS>=0.5$). It suggests that the recommendation accuracy for the hot items is almost maximum and cannot be improved anymore. However, niche items' accuracy is quite low and has plenty of room for improvement. Therefore, in order to design an more effective personalized recommendation method than current ones, it is crucial to solve the cold start problem~\cite{EPL28002}, i.e. to improve the recommendation for niche items. Another interesting finding is that only the ranking scores for unpopular items are affected by the reshuffling process while the ranking score for popular items stays almost the same. Since lots of items are with low degree in the sparse networks such as Delicious and Amazon, the average ranking score increases with the reshuffling process. In the Movielens and Netflix networks where the links are relatively dense, fewer items are with low degree in these networks. Accordingly, the average ranking scores for MD and CF do not increase much. From the practical point of view, if one want to enhance the robustness of the recommender system, the most effective way is to preserve the recommendation result for niche items since they are sensitive to randomness.

\begin{figure}
  \center
 \includegraphics[width=9cm]{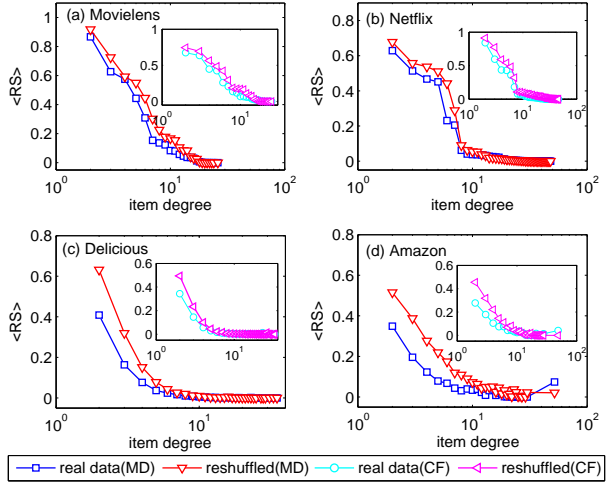}\\
  \caption{(Color online) Dependence of ranking score $<RS>$ on the item degree. The $<RS>$ is averaged by the same process in fig.2. The main figures are the results of MD method while the inserts are the results of CF method.}
\end{figure}

Precious study reveals that hybrid of the MD and HC methods can result in significant improvement in both recommendation accuracy and item diversity~\cite{PNAS4511}. Actually, this hybrid method is implementable because the HC method can effectively catch the users' taste for niche items. As the recommendation accuracy for HC method in the reshuffled networks is almost the same as random recommendation ($<RS>=0.5$), the hybrid method is impossible to be carried out in the systems where users randomly choose their items. It means that users' behavior patterns in real systems are essential for solving the diversity-accuracy dilemma of recommender systems.

\section{Conclusion}

The development for network science has greatly improved the function as well as our understanding to many real systems. In recommendation which is considered as a promising way to solve the problem of information overabundance, researchers have designed the network-based inference methods to improve the recommendation performance. For example, with the help of some typical physics dynamics on the bipartite networks, the mass diffusion and heat conduction algorithms have been proposed to improve the recommendation accuracy and diversity respectively.

In this paper, we investigate the users' online behavior patterns and related effect on information filtering. we compare the real bipartite networks with the reshuffled networks in which users' behavior patterns are gradually removed. we find that online users have significant taste diversity and their interests for niche items highly overlap. In addition, we find that the performance of popularity-based recommendation methods don't rely on the real network structure while the performance of personalized recommendation methods is strongly related to it. Detail study on the personalized methods indicates that recommendation accuracy for hot items is almost maximum and quite robust to the reshuffling process. On the contrary, niche items cannot be accurately recommended without real users' behavior properties.

From the practical point of view, in order to design a more accurate personalized recommendation method than current ones, our results suggest that it is crucial to improve the recommendation for niche items. If one wants to enhance the robustness of the recommender system, the most effective way is to preserve the recommendation result for niche items. Therefore, our work may shed some light for developing a new recommender system with both higher accuracy and better reliability.

\section*{Acknowledgement}
We thank Y.-C. Zhang, M. Medo and C. H. Yeung for the useful suggestions. This work is
supported by the Swiss National Science Foundation under Grant No.
(200020-121848).

\end{document}